\documentclass[aps,pre,preprint,superscriptaddress,showpacs]{revtex4}%
\usepackage{amssymb}
\usepackage[dvips]{graphicx}
\usepackage{colordvi}
\usepackage{times}

\newcommand{\bea}{\begin{eqnarray}}
\newcommand{\eea}{\end{eqnarray}}
\newcommand{\be}{\begin{equation}}
\newcommand{\ee}{\end{equation}}
\newcommand{\eqref}[1]{(\ref{#1})}

\newcommand{\id}{\textrm{d}}

\begin{document}

\title{A meaningful expansion around detailed balance}

\author{Matteo Colangeli}\affiliation{Dipartimento di Matematica, Politecnico di Torino, Corso Duca degli Abruzzi 24, 10129 Torino, Italy}\email{colangeli@calvino.polito.it} \author{Christian Maes} \affiliation{Instituut voor Theoretische Fysica, K. U. Leuven, 3001 Leuven, Belgium} \author{Bram Wynants} \affiliation{Institut de Physique Th\'eorique, CEA-Saclay, F-91191 Gif-sur-Yvette Cedex, France}


\begin{abstract}
We consider Markovian dynamics modeling open mesoscopic systems which are driven away from detailed balance by a nonconservative
force.
A systematic expansion is obtained of the stationary distribution around an equilibrium
reference, in orders of the nonequilibrium forcing.
The first order around equilibrium has been known since the work of McLennan (1959), and involves the transient irreversible entropy flux. The expansion generalizes the McLennan formula to higher orders, complementing the entropy flux with the dynamical activity.  The latter is more kinetic than thermodynamic and is a possible realization of Landauer's insight (1975) that, for nonequilibrium, the relative occupation of states also depends on the noise along possible escape routes. In that way nonlinear response around equilibrium can be meaningfully discussed in terms of two main quantities only, the entropy flux and the dynamical activity.  The expansion makes mathematical sense as shown in the simplest cases from exponential ergodicity.

\end{abstract}

\pacs{05.40.-a, 05.70.Ln, 05.10.Gg}


\maketitle

\section{Introduction}
Recent years have seen an intensive search for a theoretical framework underlying nonequilibrium fluctuations.  The nature of nonequilibrium
is diverse and rich but we hope to uncover some unifying structures that, ideally, would give an extension of the Gibbs formalism.  Still, today, the mathematical models are very often simplified to Markov dynamics for jump or diffusion processes that represent open systems in weak contact with one or more reservoirs. These systems (such as molecular configurations) can be very small but nevertheless we have learnt how to attach thermodynamic meaning to various path-dependent quantities, see e.g.~\cite{sek,mprf}, as they relate to what happens in the large environment.  Most of that has concentrated around the concepts of energy and entropy, so that much of standard irreversible thermodynamics also got a formulation for mesoscopic systems.  Another question has concerned response theory for these Markov dynamics and some systematics have been obtained there also, e.g. in \cite{ag,el,lip,mar,orl,vil}.  A related issue concerns the characterization of the stationary distribution for a dynamics that breaks the condition of detailed balance.  Of course we need a physically meaningful violation of detailed balance, and a useful interpretation is given by the condition of local (sometimes called, generalized) detailed balance, \cite{LB,leb,kls,der,spo}.  The latter requires that the power dissipated to the environment during a transition equals the logarithmic ratio between forward and backward transition rates. From there one can hope to find the corresponding stationary distribution in terms of these irreversible entropy fluxes. That is exactly what was achieved by McLennan in 1959, \cite{mcl59}.  In that way, the stationary distribution picks up thermodynamic information and is no longer ``just a stationary solution'' of the Master equation.  However, that McLennan proposal only works close-to-equilibrium \cite{jmp,KN,sa,ha}.  There does not appear to be a ready extension beyond the linear regime in terms of entropy considerations only.  The present paper takes a next step but we need to go beyond the purely entropic concepts we are used to from heterogeneous equilibrium.  That is, to go to second and higher order in an expansion around detailed balance we need another concept to complement the entropy fluxes.  That novel quantity is called the dynamical activity and is much related to the notion of escape rate: it measures the reactivity and instability of a trajectory. Dynamical activity is thus much more concerned with kinetics than it is embedded into thermodynamics but by introducing it, we can complete the expansion beyond linear order in the nonconservative forces around equilibrium.  In that sense, we add to the spirit of the McLennan proposal the insights of Landauer and others that for nonequilibria, the noise behavior along in- and outgoing trajectories enters critically into the determination of state plausibilities, \cite{land1,land2,land}.\\

In the expansion of the stationary distribution, every term at any order in the nonequilibrium forcing, just contains the same (dynamical) observables in various combinations of time-correlation functions under the reference equilibrium process.  For example, in some precise sense correct up to third order we get
\[
\rho(x) =\rho_o(x)\,[ 1 - \langle S\rangle_x^o + \frac 1{2}\,\langle S \,\mathcal{T}_1\rangle_x^o-\frac{1}{24}\langle S \,(S^{2}+3\mathcal{T}^{2}_1-12 \mathcal{T}_2)\rangle_x^o ]
\]
for the stationary distribution $\rho$ on states $x$ in terms of the equilibrium distribution $\rho_o$.  The averages $\langle \cdot \rangle_x^o$ are over the detailed balance process started from state $x$ while $S$ is the irreversible entropy flux and $\mathcal{T}_1, \mathcal{T}_2$ denote the first and second order to the dynamical activity.  These path observables $S$ and $\mathcal{T}_{1,2}$ depend of course on the nonequilibrium dynamics: $S$ is basically determined by the work done by the nonconservative force in a particular trajectory, and $\mathcal{T}_{1,2}$ measures the expected dynamical activity long the followed trajectory.  Specific details, rewriting and mathematical precision follow below.\\

In the next section, we specify the Markovian set-up and we define the various objects such as $\rho, S$ and $\mathcal{T}$ in the above.  We concentrate on
overdamped diffusions and jump processes for introducing (in a particular way) the nonequilibrium driving.  Sections \ref{mai}--\ref{gene}--\ref{det} give the main idea and the structure of the expansion with some writing out for specific models to lowest order.  The last section about nonlinear response suggests some immediate application before we conclude the paper. Appendix B defines the entropy flux and the dynamical activity for  underdamped diffusions.

\section{Set-up of Markov stochastic dynamics}\label{mar}

We restrict our analysis to Markovian stochastic models for mesoscopic systems driven by a nonconservative force, i.e., a force that cannot
be derived from a potential. We imagine such systems to be immersed in an environment
in thermal equilibrium at some inverse temperature $\beta$.  All conditions are time-independent and we want to characterize the statistical
distribution of states for our open system when reaching stationarity. \\
A first standard choice is to consider overdamped diffusions for state $x\in {\mathbb R}^n$, according to which
\be
\dot{x}_t = \chi \cdot F(x_t) + \sqrt{2\frac{\chi}{\beta}}\,\xi_t,\qquad \xi_t=\mbox{ standard white noise}\label{overd}
\ee
where the mobility $\chi$ is a positive definite $n\times n$-matrix not depending on $x$ (for simplicity only).  The total force equals
\[
F = \epsilon f - \nabla U \label{force}
\]
for the energy $U$  of the system, and with $f$ the nonconservative force with amplitude $\epsilon$ (our small number).
When $\epsilon = 0$, the dynamics satisfies the condition of detailed balance, and given enough time
the distribution of states converges to the equilibrium distribution $\rho_o(x) \propto e^{-\beta U(x)}$, so we assume.\\
Whenever $\epsilon \neq 0$ the system is not in equilibrium. The Fokker-Planck equation
for the evolution of distributions $\mu_t$ is
\[ \frac{\partial \mu_t}{\partial t}(x) + \nabla\cdot [\chi F(x)\mu_t(x)
                                            -\frac{\chi}{\beta}\nabla\mu_t(x)] = 0  \label{FP} \]
The stationary solution $\rho$ thus satisfies $\beta \nabla\cdot F(x)\rho(x) = \Delta \rho(x)$,
but we would be much helped by further more explicit, physical or systematic understanding of that $\rho$.  One option is to find an expansion for
that stationary distribution $\rho(x)$ of the system in orders of $\epsilon$, assuming uniformly and exponentially fast relaxation behavior.  That is the programme of the present paper and we will find that the main quantities in such an expansion are directly related to two specific path observables, that we now introduce.\\
We fix a large time-interval $[0,T]$ and we consider paths $\omega = (x_t, t\in [0,T])$, on which we define
\begin{eqnarray}\label{tau}
 \mathcal{T}(\omega) &=& \frac{\epsilon^2\beta}{2}\int_0^T \id t f\cdot \chi f - \epsilon\beta\int_0^T \id t f\cdot \chi\nabla U + \epsilon \int_0^T \id t \chi\nabla\cdot f\\
 S(\omega) &=& \epsilon\beta\int_0^T \id x_t \circ f(x_t)\nonumber
\end{eqnarray}
The last stochastic integral (with the $\circ$) is in the sense of Stratonovich; in that way $S$ is identified with $\beta$ times the work done by the nonconservative
force --- for short, we speak about the entropy flux $S$.  The quantity $\mathcal{T}$ is less familiar, and it contains both order $\epsilon$ and order $\epsilon^2$.  Correspondingly, we write $\mathcal{T} = \mathcal{T}_1 + \mathcal{T}_2$ for the first and second order.  Its meaning is best understood before the continuum limit, in terms of jump processes to which we turn next.\\
We consider a Markov jump process with discrete states $x$ and as a reference we take jump rates $k_o(x,y)$ for the transition $x\rightarrow y$ which satisfy detailed balance, i.e.,
\[
 k_o(x,y)\, e^{-\beta U(x)} =  e^{-\beta U(y)}\,k_o(y,x)
\]
In this case the stationary Master equation is solved by $\rho_o(x) \propto \exp(-\beta U(x))$.
To this reference we add an extra flux $f(x,y) = -f(y,x)$ of energy in the transition $x\rightarrow y$, and write
\begin{equation}\label{psi}
  k_{\epsilon}(x,y) = k_o(x,y)\,e^{\frac{\beta\epsilon}{2} f(x,y)}
  \end{equation}
To establish nonequilibrium we ask that the fluxes $f(x,y)$ cannot all be rewritten as the difference $V(x)-V(y)$ of a unique potential $V$, which means that there are loops
$x_1\rightarrow x_2\rightarrow\ldots x_n=x_1$ over which the sum $f(x_1,x_2) + f(x_2,x_3) +\ldots f(x_{n-1},x_n) \neq 0$ does not vanish.
In terms of the condition of local detailed balance, these fluxes $f(x,y)$ should be interpreted as the product of a displacement of a certain quantity and a nonconservative force, see~\cite{kls}, but here we do not need this formulation. The $\epsilon$ is the magnitude of the nonequilibrium forcing.
Finally, in \eqref{psi} we have chosen to omit an extra symmetric prefactor $\psi_{\epsilon}(x,y)$ because the more general choice for \eqref{psi} would be $  k_{\epsilon}(x,y) / k_o(x,y) = \psi_{\epsilon}(x,y)\,\exp[\beta\epsilon f(x,y)/2]$.  We take however $\psi_{\epsilon}(x,y) = \psi_{\epsilon}(y,x) =1$ for much greater simplicity.\\
  Again, for $\epsilon\neq 0$, the stationary probability law $\rho$ is only known indirectly as solution of the Master equation
  $\sum_{y}[\rho(y)k(y,x) - \rho(x)k(x,y)] =0$ for all $x$.\\
This time the entropy flux is
\begin{equation}\label{sstar} S(\omega) = \epsilon\beta\,\sum_{t\leq T}f(x_{t^-},x_t)
\end{equation}
as a sum over the jump times $t$ in the trajectory $\omega = (x_s, s\in [0,T])$, and the dynamical activity is
\begin{equation}\label{tauju}
\mathcal{T}(\omega) = 2\int_0^T \id t \sum_{y}k_o(x_t,y)[e^{\frac{\beta\epsilon}{2} f(x_t,y)}-1]
\end{equation}
Now we see better where the name activity comes from: $\mathcal{T}(\omega)$ is the difference in the escape rates, integrated over the trajectory.
The escape rate $\sum_y k(x,y)$ measures the frequency by which the system exits state $x$, and in that way it counts the expected number of transitions away from $x$.  In other words, $\mathcal{T}(\omega)$ sees how the escape rate away from the trajectory $\omega$ changes when adding the forcing $f$.  In the appropriate rescaling the expression
\eqref{tau} is simply the continuum limit of \eqref{tauju} from Markov jump to (overdamped) diffusions.  We do that computation in Appendix A; the case of underdamped or inertial diffusions is shortly discussed in \ref{inert}.  We repeat that we use here \eqref{psi}, i.e., with prefactor $\psi_\epsilon=1$; otherwise the expression for the activity gets more complicated --- the major part of the analysis would however remain unchanged.

\section{Expansion: main idea}\label{mai}
The expansion of the stationary distribution starts from a simple idea which was applied already in \cite{gklm,lms,KN,jmp}: the single-time distributions on states, and in particular the stationary distribution, can be obtained from its embedding in the path space distribution.  The latter is the distribution on the level of trajectories or paths $\omega$ and gives the weight $P(\omega)$ for path-integrals.  $P$ is much more directly obtained and is much better-behaved than its projections on single time layers.  In fact, we can give explicit expressions for the ``action'' $A(\omega)$ in
\begin{equation}\label{act}
P(\omega) =e^{-A(\omega)}\,P^o(\omega)
\end{equation}
that connects the distribution $P$ on paths starting from $\rho_o$ but {\it with} driving $f$, with the full equilibrium reference distribution $P^o$, see e.g. \cite{j2} for some useful techniques. The action $A$ is typically local in space-time and thus is similar to Hamiltonians or Lagrangians that we meet in (equilibrium statistical) mechanics, see e.g. \cite{poincare,mprf}.
We can verify that
\[
A = (\mathcal{T} - S)/2
\]
as defined above for overdamped and jump processes.  Furthermore, the action $A$ in \eqref{act} is left unchanged when both processes start from the same state $x$,
\begin{equation}\label{act1}
P_x(\omega) =e^{-A(\omega)}\,P^o_x(\omega)
\end{equation}

Now comes the embedding.  As defined before, both nonequilibrium and equilibrium processes in \eqref{act} start at time $t=0$ from data distributed with the equilibrium $\rho_o$.  At time $T$ later, the probability to find the driven system in state $x$ is
\[
 p(x,T) = \left<\delta(x_T-x)\right>^{\epsilon}_{\rho_0}
\]
averaging over the trajectories for the nonequilibrium dynamics.
We assume that for $T\uparrow +\infty$ there is exponentially fast convergence  $p(x,T) \rightarrow \rho(x)$  (in the sense of densities) to the stationary distribution of the nonequilibrium process, uniformly in $\epsilon$. By \eqref{act} we can rewrite
\begin{equation}\label{prad} p(x,T) = \left<\delta(x_T-x)\,e^{-A(\omega)}\right>^o \end{equation}
which is now an expectation value for the full equilibrium (detailed balance) process (and then we omit the subscript $\rho_o$).
 By time-reversal invariance, \eqref{prad} equals
\begin{equation}\label{almost} p(x,T) = \left<\delta(x_0-\pi x)\,e^{-A(\theta\omega)}\right>^o
          = \rho_o(x)\,\left<e^{-A(\theta\omega)}\right>^o_{x_0 = \pi x}
\end{equation}
where the time-reversal operator $\theta$ acts as
\[
 \theta \omega = ((\pi x)_{T-t}, 0\leq t\leq T)
\]
with $\pi{x}$ equal to $x$ except for flipping the velocities (if they are part of the state-description)
or other variables with negative parity under time-reversal.  In equilibrium $\rho_o(\pi x) = \rho_o(x)$.
Equality \eqref{almost} can still be rewritten in terms of $S$ and $\mathcal{T}$ getting
\begin{equation}\label{central}
 p(x,T) \rho_o(x)\,\left< e^{-(S+{\mathcal{T}})/2} \right>^o_{x_0 = \pi x}
\end{equation}
Indeed, the decomposition $A = (\mathcal{T} - S)/2$ follows the symmetry under time-reversal,
\begin{eqnarray*}
 S(\omega)           &=& A(\theta\omega) - A(\omega)\\
 \mathcal{T}(\omega) &=& A(\theta\omega) + A(\omega)
\end{eqnarray*}
The fact that the quantity $A(\theta\omega) - A(\omega)$ {\it is} the excess entropy flux from the system
into the environment during the process $\omega$ is one of the main discoveries for the construction
of nonequilibrium statistical mechanics of the last decade, see \cite{mn,crooks} and the references in e.g.~\cite{vol,mar,poincare}.
Excess means the difference between the nonequilibrium process and the reference equilibrium
process. Specifically, this excess is here equal to the work done
by the nonequilibrium force $\epsilon f$, multiplied by $\beta$.
The dynamical activity $\mathcal{T}$ has been introduced and used before \cite{lec,maarten}
but has no thermodynamic tradition.  Its role is kinetic and that it influences the relative stability of states was somehow emphasized long before, cf.~\cite{land1,land2,land}.\\

The left-hand side of \eqref{central} is assumed to converge exponentially fast to the stationary law $\rho(x)$, but there is a problem with its right-hand side because
 both $S$ and ${\mathcal{T}}$ are time-extensive of order $T$, being sent to infinity.  We will therefore need to control the limit $T\uparrow +\infty$ and to worry about the exchange with the sum of the perturbation series.  An important point here that follows from \eqref{act1}, is the normalization
\begin{equation}\label{norma}
 \left< e^{(S-{\mathcal{T}})/2} \right>^o_x = 1
\end{equation}
valid under the equilibrium process but started from an arbitrary state $x$. That itself can be expanded in orders of $\epsilon$, and takes care of many cancellations.  Finally, one must use that $S$ is anti-symmetric, and $\mathcal{T}$ is symmetric under time-reversal, so that over all time-intervals $[0,T]$
\begin{equation}\label{asy}
\left< S^n\,{\mathcal{T}}^m\right>^o = 0,\quad n \mbox{ odd}
 \end{equation}
 for all $m$ and for all odd powers $n$.  With these ingredients \eqref{central}--\eqref{norma}--\eqref{asy}, combined with fast relaxation for the equilibrium process, all is in place to start a systematic expansion. In the particular cases we have in mind, see Section \ref{mar}, the path function $S$ is simply first order in $\epsilon$ and $\mathcal{T}$ is either $\mathcal{T} = \mathcal{T}_1 + \mathcal{T}_2$ second order in $\epsilon$ for diffusions, see \eqref{tau}, or of arbitrary order $\mathcal{T} = \mathcal{T}_1 + \mathcal{T}_2 + ...$ for Markov jump processes.  For practical matters our expansion including second or third order around equilibrium is already new and relevant.

\section{General expansion}\label{gene}

Formally, expanding (\ref{central}) just gives

\[
\frac{p(x,T)}{\rho_o(x)}=1 -\frac{1}{2}\left<S + \mathcal{T}_1\right>^o_{x_0 = \pi x}
 +\frac{1}{2}\left<-\mathcal{T}_2 +
\frac{(S+\mathcal{T}_1)^2}{4}\right>^o_{x_0 = \pi x} + O(\epsilon^3)
\]
On the other hand, \eqref{norma} gives
\[ 0 = \frac{1}{2}\langle
S-\mathcal{T}_1\rangle^{o}_{x_{0}=\pi x},\;\,
0 = \langle \frac{(S-\mathcal{T}_1)^2}{8} -  \frac{\mathcal{T}_2}{2}\rangle^{o}_{x_{0}=\pi x}, \ldots\]
Adding or subtracting these relations from the corresponding orders of the expansion of the stationary distribution
simplifies matters.  In the end, in every order of the expansion we can choose that only those terms  survive
which are averages of quantities antisymmetric in time,
\begin{eqnarray}\label{loworder2}
\frac{p(x,T)}{\rho_o(x)} =&& 1 -\left<S\right>^o_{x_0 = \pi x}
+\frac{1}{2}\left<S\mathcal{T}_1\right>^o_{x_0 = \pi x}\\
&&-\frac{1}{24}\left<S( S^{2}+3\mathcal{T}_1^{2}-12\mathcal{T}_2)\right>^o_{x_0 = \pi x} + O(\epsilon^4) \nonumber
\end{eqnarray}
The above considerations can be systematized. The formal expansion that results, after also taking into account the normalization in (\ref{norma}), is
\be
p(x,T) = \rho_o(x)\left[1+ \sum_{m=1}^{\infty}\Big<B_{m}\left(-\frac{S+\mathcal{T}}{2}\right)
-B_{m}\left(\frac{S-\mathcal{T}}{2}\right)\Big>^{o}_{x_0=\pi x}\right] \label{exact}\ee
where we introduced a shorthand notation for the path-dependent functionals $B_{m}(G)$ acting on path observables $G(\omega,\epsilon)$, defined via
\bea
B_{m}(G) &=& B_{m}(G_1,...,G_{m-k+1}) \nonumber\\
&=&\sum_{k=1}^m\sum_{\sigma}\frac{1}{b_1!...b_{m-k+1}!}\left(\frac{G_1}{1!}\right)^{b_1}...\left(\frac{G_{m-k+1}}{(m-k+1)!}\right)^{b_{m-k+1}} \label{Bell}
\eea
for $G_\ell(\omega) = d^\ell G/d\epsilon^\ell (\omega,0)$. The sum in (\ref{Bell}) extends over all sequences $\sigma$ of non-negative coefficients $\sigma= (b_1,...,b_{m-k+1})$, such that $\sum_{j=1}^{m-k+1}b_j=k$ and $\sum_{j=1}^{m-k+1}j b_j=m$. Note that the $B_m$ are versions of the so called complete Bell polynomials.  There exist alternative, although equivalent, expressions for (\ref{exact}), see e.g. \cite{johnson}. On the other hand, the expression we use has the advantage of being  compact and suitable for numerical implementation.\\

From (\ref{exact}) and (\ref{Bell}), we can write the explicit expression for the $m$-th order in $\epsilon$.  Remember that we write $\mathcal{T}=\mathcal{T}_1 + \mathcal{T}_2 + \ldots$ and that $S$ is of order $\epsilon$ while $\mathcal{T}_n$ is of order $\epsilon^n$.
The result is
\begin{equation}
\frac{p(x,T)}{\rho_o(x)} = 1+ 2\sum_{m=1}^{\infty}\sum_{\sigma_m}(\frac{-1}{2})^{\sum b_j}\,\frac{1}{b_0!b_1!...b_{m}!}
 \Big<S^{b_0}\mathcal{T}_1^{b_1} \mathcal{T}_2^{b_2}...\mathcal{T}_{m}^{b_{m}}\Big>^{o}_{x_0=\pi x} \label{exact2}
\end{equation}
The sum in (\ref{exact2}) extends over all sequences $\sigma_m$ of non-negative integers $(b_0,b_1,...,b_{m})$, such that $b_0$ is odd and $b_0 + \sum_{j=1}^{m}\,j b_j=m$.  For illustration to construct \eqref{loworder2}, $m=1$ requires $b_0=1$ and all other $b_j=0$; $m=2$ requires $b_0=b_1=1$ with all the other $b_j=0$; $m=3$ allows three cases $b_0=b_2=1$, $b_0=1, b_1=2$ and $b_0=3$ each time with all other $b_j=0$.\\
In the case of diffusions (where $p(x,T)$ must be understood as a probability density with respect to $\id x$), see \eqref{tau}, we have $\mathcal{T}_n=0$ for $n>2$ so that we must then also require $b_j=0, j\geq 3$ in each $\sigma_m$.\\

The first important thing to observe about the expansion \eqref{exact2} is that it converges for fixed state $x$ and uniformly in time $T\uparrow \infty$ if there is $c=c(x)<\infty$ such that each term is bounded like
\begin{equation}\label{bou}
|\Big<S^{b_0}\mathcal{T}_1^{b_1} \mathcal{T}_2^{b_2}...\mathcal{T}_{m}^{b_{m}}\Big>^{o}_{x_0=\pi x}| \leq \epsilon^m \, c^{\sum b_j}
\end{equation}
The reason is that
\[
\sum_{m=1}^{\infty}\epsilon^m\sum_{\sigma_m}(\frac{c}{2})^{\sum b_j}\,\frac{1}{b_0!b_1!...b_{m}!}\leq \sum_{m=1}^{\infty}\epsilon^m\sum_{k=1}^m\frac{c^k\,(m+1)^{k}}{2^k k!}
\]
converges for small enough $\epsilon$.\\
We only argue for \eqref{bou} explicitly for the first and the second order.  The first order is the McLennan formula (see immediately below) and has been treated before in~\cite{jmp}.  The second order adds a new complication (to be treated below) and that complication is repeated for the higher order terms and can be solved in the same way.  At any rate we cannot quite leave it with \eqref{loworder2} or with \eqref{exact2} because we are interested in the limit $T\uparrow \infty$ and $S,\mathcal{T}$ do not make any sense in that limit.  We thus need a further rewriting for which we need some more model-dependent input and to which we turn next.

\section{Expansion details}\label{det}
\subsection{McLennan formula: the first order}

The first order in the expansion of the stationary distribution (\ref{loworder2})
has been known for a long time \cite{mcl59} and has been reconsidered more recently in \cite{KN,jmp}.
In fact, the idea of obtaining the McLennan-formula via the embedding described under Section \ref{mai} originates from
\cite{KN}.  The way how to deal with the limiting behavior $T\uparrow \infty, \epsilon \downarrow 0$ was treated in \cite{jmp}.  We briefly repeat this and we concentrate on the Markov processes of Section \ref{mar}.\\

The main point is that
\begin{equation}\label{mcl}
\left<S(\omega)\right>^o_{x} = \epsilon\,\beta\, \int_0^T \id t \langle w(x_t) \rangle^o_x
\end{equation}
where, for jump processes $w(x) = \sum_y k_o(x,y) f(x,y)$, and for overdamped diffusions
$w = \chi\nabla \cdot f/\beta - \chi f\cdot \nabla U$, see \cite{jmp}.  The expression \eqref{mcl} allows the limit $T\uparrow \infty$ uniformly in $\epsilon$ once we
assume the equilibrium process to be irreducible and exponentially ergodic.\\

Plugging \eqref{mcl} into \eqref{exact2} thus gives the linear order expression
\begin{equation}\label{macl}
\rho(x)/\rho_o(x) = 1 - \epsilon\,\beta\,h(x) + O(\epsilon^2)
\end{equation}
with
\[
h(x) = \int_0^\infty \id t \, \langle w(x_t)\rangle_x^o
\]
in which the integral is exponentially convergent.  The relation with local equilibrium distributions is also discussed in \cite{jmp}.

\subsection{Second order}

We look at the $m=2$ term in \eqref{exact2}.
The main object to consider for jump processes is
\[
\int_0^T \id s \left< \sum_t^T f(x_{t^-},x_t)\,\sum_y k_o(x_s,y) f(x_s,y)\right>_{x}^o
\]
where the sum is over the jump times.  On the other hand, for overdamped diffusions we can
introduce $\sigma(x_t,\id x_t) =  \epsilon\beta\,\id x_t\circ f(x_t)$
and   $\mathcal{T}_1(x) = - \beta\epsilon f(x)\cdot \chi \nabla U + \epsilon\chi\nabla\cdot f$, so that we must deal then with
\begin{equation}\label{seco}
h_2(x) \equiv \int_0^T \id s \int_0^T \left<\sigma(x_t,\id x_t)\,\mathcal{T}_1(x_s)\right>_{x}^o
\end{equation}
We can keep with this ``diffusion-''notation also for the ``jump-''case if we then take $\sigma(x_t,\id x_t) =  \epsilon\beta\,\id {\cal N}_t f(x_{t^-},x_t)$ where ${\cal N}_t$ is the Poisson process counting jumps,
and   $\mathcal{T}_1(x) = \epsilon\beta \sum_y k_o(x,y) f(x,y)$.  We give the argument for the jump-case.

Because we are dealing with a correlation function of two time-extensive quantities, we have to use
twice the exponential convergence to the equilibrium expectation.
We will use the bound $|\langle f(x_u) g(x_0)\rangle^o - \langle f(x_u) \rangle^o \,\langle g(x_0)\rangle^o| \leq C_fC_g e^{-\alpha u}$, for some positive $\alpha$ and where $C_f$ and $C_g$ bound the functions $f$, respectively $g$.\\
To this end, we split the $s$-integral in \eqref{seco} over $[0,t]\cup[t,T]$.  For the first part,
\[
\int_0^T \int_{0}^t  \id s\,\left<\sigma(x_t,\id x_t)\mathcal{T}_1(x_s)\right>_{x}^o
= \int_0^T \id t \int_{0}^t  \id s\,\left<\mathcal{T}_1(x_t)\mathcal{T}_1(x_s)\right>_{x}^o
\]
as follows from \eqref{norma}.  For the second part we use
\[
\int_0^T \int_{t}^T  \id s\,\left<\sigma(x_t,\id x_t)\mathcal{T}_1(x_s)\right>_{x}^o
\int_0^T \id t\,\int_0^{t} \, \left<\sigma(x_s,\id x_s)\mathcal{T}_1(x_t)\right>_{x}^o \label{terms}
\]
so that always $s<t$ in what follows.  Therefore,
\begin{equation}\label{ers}
 h_2(x) = \int_0^T \id t \int_0^{t}  \left<[\sigma(x_s,\id x_s) + \mathcal{T}_1(x_s)\,\id s]\mathcal{T}_1(x_t)\right>_{x}^o
\end{equation}
First of all, the integrand has an equilibrium expectation equal to zero, as one can see from time-reversal invariance of equilibrium:
\begin{eqnarray*}
\left<[\sigma(x_s,\id x_s) + \mathcal{T}_1(x_s)\,\id s]\mathcal{T}_1(x_t)\right>^o = \left<[-\sigma(x_t,\id x_t) + \mathcal{T}_1(x_t)\,\id t]\mathcal{T}_1(x_s)\right>^o
\end{eqnarray*}
As $t$ is always bigger than $s$ in the integrand of (\ref{ers}), we can do the same trick as before by replacing $\left<\sigma(x_t,dx_t)\mathcal{T}_1(x_s)\right>^o$ by  $\left<\mathcal{T}_1(x_t)\id t\mathcal{T}_1(x_s)\right>^o$. This shows us indeed that the equilibrium expectation of the integrand in (\ref{ers}) is zero.
We can make that more explicit by substituting the expressions for the jump-case, to have
\begin{eqnarray}\label{po1}
&& \left<[\sigma(x_s,\id x_s) + \mathcal{T}_1(x_s)\,\id s]\mathcal{T}_1(x_t)\right>_{x}^o=\nonumber\\
 && (\epsilon\beta)^2\,\langle [\id {\cal N}_s f(x_{s^{-}},x_s) + \zeta(x_s)\id s]\langle \zeta(x_t)|x_s\rangle^o\rangle^o_x
 \end{eqnarray}
 for $\zeta(x) \equiv \sum_y k_o(x,y)f(x,y)$.
Therefore, with $p_s(x,y)$ the transition probability for the detailed balance reference dynamics to find $y$ at time $s$ when starting from $x$,
\begin{eqnarray}\label{prob}
&& \left<[\sigma(x_s,\id x_s) + \mathcal{T}_1(x_s)\,\id s]\mathcal{T}_1(x_t)\right>_{x}^o =\\
&&(\epsilon\beta)^2\,\id s\,\sum_{y,z} [p_s(x,z)\, k_o(z,y)\, f(z,y) + p_s(x,y)\,k_o(y,z)f(y,z)]\langle \zeta(x_{t-s})\rangle^o_y\nonumber
\end{eqnarray}
which indeed vanishes upon replacing $p_s(x,y)\rightarrow \rho_o(y)$ by the equilibrium distribution because
$\rho_o(z)\, k_o(z,y)\, f(z,y) + \rho_o(y)\,k_o(y,z)f(y,z) = 0$ by detailed balance.\\
Furthermore,
the equilibrium expectation of $\mathcal{T}_1$ is zero, $\langle \mathcal{T}_1(x_t)\rangle^o = 0$, so that
\[
|\langle \zeta(x_{t-s})\rangle^o_y|\leq C(y)\,e^{-\alpha(t-s)}
\]
The rest of the argument is straightforward by using that $p_s(x,y)$ is exponentially close to $\rho_o(y)$ as a function of the time $s$:
\[ |h_2(x)|
\leq \int_0^T \id t \int_0^t \id s \,e^{-\alpha (t-s)}e^{-\alpha' s}C'(x) \]
which is clearly finite as $T\uparrow \infty$.

\section{Application to nonlinear response}
One of the very first applications of the McLennan formula (the first order as in \eqref{macl}) is the derivation of linear response  around equilibrium.  It is indeed possible to derive the Green-Kubo relations from it; see e.g. Section IIIB in \cite{jmp}.  The analogue can be done also for higher order response as we now indicate.\\
We imagine the system in equilibrium up to time zero. At that time an external stimulus $\epsilon f$ is added, as modeled in the set-up of Section \ref{mar}, driving the
system away from equilibrium. We can then estimate, using \eqref{exact2} say to second order in $\epsilon$, for an observation $Q$ at time $T$,
\begin{eqnarray*}
\langle Q(x_T)\rangle^{\epsilon}_{\rho_0}
&=& \langle Q\rangle^{o} - \langle Q(\pi x_0)S(\omega)\rangle^{o}
+ \frac{1}{2}\langle Q(\pi x_0)S(\omega)\mathcal{T}_1(\omega)\rangle^{o} + o(\epsilon^2)\\
&=& \langle Q\rangle^{o} + \langle Q(x_T)S(\omega)\rangle^{o}
- \frac{1}{2}\langle Q(x_T)S(\omega)\mathcal{T}_1(\omega)\rangle^{o} + o(\epsilon^2)
\end{eqnarray*}
When the perturbation $\epsilon f$ is the difference or the gradient of a potential $\epsilon V$, then the entropy flux $S = \beta\epsilon[V(x_T)-V(x_0)]$, assuming that
$V(x) = V(\pi x)$ (see formulae (\ref{tau})
and (\ref{sstar})). The expectation value then reads, up to second order,
\begin{eqnarray}\label{fdr}
\langle Q(x_T)\rangle^{\epsilon}_{\rho_0} &=&
\langle Q\rangle^{o} + \beta\epsilon\langle Q(x_T)[V(x_T)-V(x_0)]\rangle^{o}\nonumber\\
&& - \frac{\beta^2\epsilon^2}{2}\int_0^T\id s \langle Q(x_T)[V(x_T)-V(x_0)]\,L_oV(x_s)\rangle^{o}
\end{eqnarray}
The first order term is consistent with the (equilibrium) fluctuation-dissipation theorem.
In the second order term in \eqref{fdr} we have used the expression for the backward generator
 $L_0V (x) = \sum_y k_o(x,y)[V(y)-V(x)]$ for jump processes, $L_oV(x)
= -\chi\nabla U \cdot \nabla V + \chi \Delta V/\beta$ for diffusion processes, to rewrite
\[
 \mathcal{T}_1(\omega) = \epsilon\,\beta\,\int_0^T \id s \,L_oV(x_s)
 \]
 when $f(x,y) = V(y) - V(x)$ (in the jump-case for $\psi_\epsilon(x,y)=1$ in
\eqref{psi}) or when $f(x) =\nabla V(x)$ (in the diffusion-case \eqref{tau}). As a consequence,
\[
 \left.\frac{\partial^2}{\partial\epsilon^2}\langle Q(x_T)\rangle^{\epsilon}_{\rho_0}\right|_{\epsilon=0} = - \frac{\beta^2}{2}\,\int_0^T \id s\langle [Q(x_T)-Q(\pi x_0)][V(x_T)-V(x_0)]\,L_oV(x_s)\rangle^{o}
\]
 Observe that the terms in the expansion \eqref{fdr} are not just expectations of observables at one fixed time; they are correlation functions. It is therefore natural to investigate also the perturbation expansion of correlation functions in the nonequilibrium process.  That follows most generally from expanding the exponential and the action in \eqref{act}--\eqref{act1}.  It is then straightforward to check e.g. that
\begin{equation}
 \left.\frac{\partial^2}{\partial\epsilon^2}\langle Q(x_t)\rangle^{\epsilon}_{\rho_0}\right|_{\epsilon=0}
= \beta\left.\frac{\partial}{\partial\epsilon}\langle \langle [Q(x_t)-Q(\pi x_0)][V(x_t)-V(x_0)]\rangle^{\epsilon}_{\rho_0}\right|_{\epsilon=0}
\end{equation}
The treatment of nonlinear response can be started differently, for example from applying Ward-Takahashi type identities starting from the fluctuation symmetry in the distribution of the entropy flux, see e.g. Section 10 in \cite{poincare} or more recently in \cite{ag,mar,orl,lip,vil}.  The advantage of the present treatment however is that all the terms in the expansion are explicitly expressed as correlation functions in the full equilibrium reference process.  On the other hand, we do not know how to extend our ideas to deterministic dissipative dynamics.  There  the steady state attractor has a lower dimension than the embedding space and the invariant density becomes singular. That aspect probably becomes less stringent for macroscopic systems, or, depending on the reduced description and the nature of the observables one can attempt a suitable projection on some smooth manifold. For a perturbation theory featuring a first order term producing the standard fluctuation-dissipation theorem and higher order terms giving rise to nonlinear response, see e.g.~\cite{BLMV} and~\cite{ru} for two different approaches.  For further convergence of ideas, it would be very helpful to identify the notion of dynamical activity within the thermodynamic formalism of smooth dynamical systems.  One natural guess would proceed via the escape rate formalism as in \cite{gas}.

\section{Conclusion}

Expansions for nonequilibrium mesoscopic systems can be made in various ways.  If one is interested in the stationary distribution away from detailed balance, as we are, we can try Born-type or Dyson-type perturbation expansions starting from the Master equation.  These can certainly be computationally useful, but that however is not the main point of this present paper.  What was attempted here is to give an expansion in terms of few (basically two) path-dependent physical observables, the entropy flux and the dynamical activity.  It is in the same spirit of many attempts dating already from the 1950-60's where a major issue was to
understand whether the stationary nonequilibrium distribution
could be described in terms of macroscopic parameters only, \cite{leb,LB,mcl59}. Here we deal with mesoscopic systems and we incorporate more recent ideas concerning the importance of the kinetics (noise and activity) for finding the relative probability of states, \cite{land}.  The utility of the expansion depends on the relevance of nonlinear response around equilibrium for stochastic Markovian dynamics, and whether natural phenomena obey the various conditions imposed.\\

The starting point of our analysis is a path-integral formulation for the distribution of histories in terms of a reference equilibrium distribution.
The breaking of detailed balance introduces irreversible fluxes, making up the time anti-symmetric term in the action, and also creates excesses in dynamical activity, governing the time-symmetric fluctuations.  Both fluxes and activities can be written down explicitly as function of the system's trajectory for Markov jump and diffusion processes, where the physical interpretation is supported by the condition of local detailed balance.
The subsequent expansion then mimics a high-temperature expansion for Gibbs distributions but now on space-time, where the exponential function gets Taylor-expanded.  Additional symmetries simplify the expansion and fast relaxation properties enable full control over the terms in the expansion, as explicitly demonstrated for the first few.  It remains however an interesting mathematical challenge to rigorously prove the convergence of the complete perturbation series under an appropriate minimal set of conditions on the dynamics.

\section{Acknowledgements}
M.C.\ acknowledges support from the Swiss National Science Foundation (SNSF).
C.M.\ benefits from the Belgian Interuniversity
Attraction Poles Programme P6/02.
 B.W.\ was junior research
fellow at the FWO, Flanders
until 30-09-2010.

\appendix

\section{Continuum limit of dynamical activity}
There is a large literature concerning the derivation of Langevin dynamics or the Fokker-Planck equation from the Master equation for jump processes.  Most of the time one develops around some small parameter like the inverse volume or the inverse number of particles in what is known as the Kramers-Moyal expansion, \cite{vk}.  We do not repeat that here, but since the dynamical activity is still relatively unknown we show how to obtain \eqref{tau} from \eqref{tauju} in a continuum limit.\\

Consider a small mesh $\delta>0$ and a walker on a ring with detailed balance transition rates
\[
k_o(x,x\pm\delta) = D\,\exp-\frac{\beta}{2}[U(x\pm \delta) - U(x)],\;\;D>0
\]
and with driven rates
\[
k(x,x\pm\delta) = k_o(x,x\pm\delta)\,\exp\frac{\beta \epsilon \delta}{2}f(x,x\pm\delta)
\]
The excess in dynamical activity when in state $x$ is then, following \eqref{tauju},
\[
k(x,x+\delta)-k_o(x,x+\delta) + k(x,x-\delta) - k_o(x,x-\delta)
\]
which we expand to order $\delta^2$ as
\begin{eqnarray}
&& D\,(1-\frac{\beta \delta}{2} U'(x))\,\big[\frac{\beta \epsilon \delta}{2}\,f(x,x+\delta) + \frac{\beta^2 \epsilon^2 \delta^2}{8} f^2(x,x+\delta)\big] \nonumber\\
+&& D\,(1+\frac{\beta \delta}{2} U'(x))\,\big[\frac{\beta \epsilon \delta}{2}\,f(x,x-\delta) + \frac{\beta^2 \epsilon^2 \delta^2}{8} f^2(x,x-\delta)\big] \nonumber\\
=&& \frac{D\beta^2\epsilon^2 \delta^2}{8}\,[f^2(x,x+\delta) + f^2(x,x-\delta)] + \frac{D\beta \epsilon \delta}{2}\,[f(x,x+\delta) - f(x-\delta,x)] \nonumber\\
&-& \frac{D\beta^2 \delta^2\epsilon}{4}U'(x)\,[f(x,x+\delta) + f(x-\delta,x)]\nonumber
\end{eqnarray}
In other words, setting also $D\beta=\chi$ and $f(x,x)=f(x)$,
\bea
&&\lim_{\delta\downarrow 0} \frac{2}{\delta^{2}}[k(x,x+\delta)-k_o(x,x+\delta) + k(x,x-\delta) - k_o(x,x-\delta)]=\nonumber \\
=&&\frac{\chi \beta}{2} \epsilon^2 f^2(x) + \chi\epsilon f'(x) -\chi\beta\epsilon f(x) U'(x)
\eea
in which we recognize the instantaneous excess in dynamical activity \eqref{tau}.

\section{Underdamped diffusion processes}\label{inert}

We consider here a Langevin dynamics for a particle with mass $m$, position $q_t$ and velocity $v_t$:
\begin{eqnarray}
 dq_t &=& v_t \id t \nonumber\\
 mdv_t &=& [F(q_t)- m\gamma v_t]\id t + \sqrt{2D}dB_t\label{underd}
\end{eqnarray}
$\gamma$ is the friction coefficient and $B_t$ is a standard Wiener process
giving rise to Gaussian white noise. The (symmetric) matrix $D$ governs the variance of that noise.
The Einstein relation between $\gamma$ and $D$ brings in the inverse temperature: $\gamma = \beta D$.
The corresponding Fokker-Planck equation for the evolution of distributions $\mu_t$ is
\begin{equation}\label{fk}
 \frac{\id }{\id t}\mu_t + \nabla \cdot J_{\mu_t} =0
 \end{equation}
for $\nabla = (\nabla_q,\nabla_v)$ and for current $J_\mu = (J_\mu^q,J_\mu^v)$ with
 \begin{equation}\label{curre}
 J_\mu^q= mv\mu,\qquad J_\mu^v = F\mu - \gamma \,mv \,\mu - D\,\nabla_v\mu
 \end{equation}
to be understood with diagonal matrices $\gamma$ and $D$.
As before we write the force $F = \epsilon f - \nabla_q U$. The equilibrium case
$\epsilon = 0$ has the equilibrium distribution
\[ \rho_o(q,v) \propto e^{-\beta[m\frac{v^2}{2}+U(x)]} \]
Here the action is
\begin{eqnarray*}
  A(\omega) &=& -\log\frac{d\mathcal{P}^{\epsilon}(\omega)}{d\mathcal{P}^0(\omega)}\\
            &=& \mathcal{T}(\omega) - S(\omega)
\end{eqnarray*}
with
\begin{eqnarray}\label{smo}
 \mathcal{T}(\omega) &=& \frac{\epsilon^2}{2}\int_0^T \id t f\cdot D^{-1} f
                         - \epsilon\int_0^T \id t f\cdot D^{-1}\nabla U - m\,\epsilon \int_0^T dv_t \circ D^{-1} f\\
 S(\omega) &=& \epsilon\beta\int_0^T \id t v \cdot f
\end{eqnarray}
One checks that $S$ equals the work done by the nonconservative force $f$, times $\beta$.
The dynamical activity $\mathcal{T}$ consists of several terms with less obvious physical meanings.
Still, it consists of in principle measurable quantities: forces on the one hand and $D$ on the other hand, which
depends on the friction coefficient and the mass of the particle.  This dynamical activity turns up in the linear response around nonequilibrium, see \cite{el}.  A natural continuation of Appendix A would be to understand the Smoluchowski limit of \eqref{smo}. We don't do that here but the expansion of Section \ref{gene} and the formula \eqref{exact2} remain unchanged.

\end{document}